\documentclass[twocolumn,showpacs,floatfix,superscriptaddress]{revtex4}
\usepackage{graphicx}
\usepackage{bm}

\begin{document}

\title{Finite temperature excitations of a trapped Bose-Fermi mixture}
\author{Xia-Ji Liu}
\affiliation{\ LENS, Universit\`{a} di Firenze, Via Nello Carrara
1, 50019 Sesto Fiorentino, Italy}
\affiliation{\ Institute of Theoretical Physics, Academia Sinica, \\
Beijing 100080, China}
\author{Hui Hu}
\affiliation{\ Abdus Salam International Center for Theoretical
Physics, P. O. Box 586, Trieste 34100, Italy}
\date{\today}

\begin{abstract}
We present a detailed study of the low-lying collective
excitations of a spherically trapped Bose-Fermi mixture at finite
temperature in the collisionless regime. The excitation
frequencies of the condensate are calculated self-consistently
using the static Hartree-Fock-Bogoliubov theory within the Popov
approximation. The frequency shifts and damping rates due to the
coupled dynamics of the condensate, noncondensate, and degenerate
Fermi gas are also taken into account by means of the random phase
approximation and linear response theory. In our treatment, the
dipole excitation remains close to the bare trapping frequency for
all temperatures considered, and thus is consistent with the
generalized Kohn theorem. We discuss in some detail the behavior
of monopole and quadrupole excitations as a function of the
Bose-Fermi coupling. At nonzero temperatures we find that, as the
mixture moves towards spatial separation with increasing
Bose-Fermi coupling, the damping rate of the monopole (quadrupole)
excitation increases (decreases). This provides us a useful
signature to identify the phase transition of spatial separation.
\end{abstract}

\pacs{PACS numbers:03.75.Kk, 03.75.Ss, 67.60.-g, 67.40.Db}
\maketitle

\section{introduction}

The impressive experimental achievement of Bose-Einstein condensation (BEC)
in the bosonic systems $^{87}${\rm Rb} \cite{jila}, $^{23}${\rm Na} \cite
{mit}, and $^7${\rm Li} \cite{rice} has initiated and stimulated a whole new
field of research in the physics of quantum atomic gases. Recently, several
groups have extended these experiments to the case of trapped Bose-Fermi
mixtures, in order to employ the ``sympathetic cooling'' to reach the regime
of quantum degeneracy for the Fermi gas. As a first step, stable
Bose-Einstein condensates immersed in a degenerate Fermi gas have been
realized with $^7${\rm Li} in $^6${\rm Li} \cite{lili}, $^{23}${\rm Na} in $%
^6${\rm Li} \cite{nali}, and very recently with $^{87}${\rm Rb} in $^{40}$%
{\rm K} \cite{lens}. In the possible next step, investigations of the
thermodynamics, collective many-body effects and other properties could be
available soon in these systems. Especially interesting is the behavior of
low-energy collective excitations, since the high accuracy of frequency
measurements and the sensitivity of collective phenomena to the interatomic
interaction make them good candidates to unravel the dynamical correlation
of the many-body system.

On the theoretical side, several analyses have been presented for
low-lying collective excitations of a trapped Bose-Fermi mixture.
Collective modes in the collisionless limit, where the collision
rate is small compared with the frequencies of particle motion in
traps, have been considered by the sum-rule approach
\cite{sumrules}, by the scaling theory \cite{liu}, or in the
random-phase approximation \cite{capuzzi,rpa}. In the
collision-dominated regime collective oscillations have been
discussed by Tosi {\it et al}. in Refs. \cite{tosi00,tosi03}, and
by the authors in Ref. \cite{liu}. These investigations have
mainly concentrated at {\em zero} temperature using the standard
two-fluid model for the condensate and the degenerate Fermi gas.
However, the realistic experiment is most likely carried out at
relatively {\em higher} temperatures, where the condensate
oscillates in the presence of a considerably large fraction of
above-condensate atoms. It thus seems timely to develop an
extension of these theories to the finite temperature.

In the present paper we investigate the low-lying collective
excitations of a spherically trapped Bose-Fermi mixture at {\em
finite} temperature in the {\em collisionless} regime. We confine
ourselves to the collective modes of the condensate, {\it i.e.},
the density oscillations of the condensate. We first calculate the
mode frequencies by using the simplest temperature-dependent
mean-field theory --- the Popov version of the
Hartree-Fock-Bogoliubov (HFB) theory --- that has been generalized
by us to a trapped Bose-Fermi mixture to study its thermodynamics
\cite{hu}. For a purely Bose gas, it is well known that the
HFB-Popov theory includes only the {\em static} mean-field effects
of the noncondensate atoms \cite{griffin} and thus predicts the
correct mode frequency only at temperatures $T\lesssim
T_0=0.65T_c$ \cite{burnett}, where $T_c$ is the critical
temperature of BEC. Above $T_0$ the noncondensate component
becomes considerably large and its dynamics should be treated on
an equal footing with that of the condensate
\cite{giorgini,stoof,zgn,fedichev,reidl}. In our case of
Bose-Fermi mixtures, the situation is more crucial. Due to the
large number of fermions, the coupled dynamics of the condensate,
noncondensate, and degenerate Fermi gas has to be taken into
account even at zero temperature. In this paper, we shall treat it
perturbatively in the spirit of the random phase approximation
(RPA) and linear response theory. We derive the explicit
expression for the frequency shift and damping rate arising from
the coupled dynamics, which in the absence of the Bose-Fermi
interaction coincides with the finding of Ref. \cite{giorgini}.
Based on this expression and the static HFB-Popov theory, we
present a detailed numerical study of the monopole and quadrupole
condensate oscillations against the Bose-Fermi coupling. The
dipole excitation is also studied and found to be consistent with
the generalized Kohn theorem.

The paper is organized as follows. In the next section we derive the theory
used in this paper. In Sec. III we apply this theory to mixtures of $^{41}%
{\rm K}-^{40}${\rm K} and $^{87}{\rm Rb}-^{40}${\rm K}, and calculate the
dispersion relation of the monopole and quadrupole excitations as a function
of the Bose-Fermi coupling. The behavior of monopole and quadrupole modes
against temperature is also discussed in detail. Finally, section IV is
devoted to conclusions.

\section{formulation}

In this section we first generalize a time-dependent mean-field scheme
developed by Giorgini \cite{giorgini} to Bose-Fermi mixtures, and derive the
equation for the small-amplitude oscillations of the condensate. Since the
formalism of this time-dependent mean-field approximation for an
inhomogeneous interacting Bose gas has already been presented in detail in
Ref. \cite{giorgini}, here we shall merely concentrate on the key points,
and indicate the necessary modification in the presence of the fermionic
component. By means of the RPA and linear response theory, we further
consider the fluctuations of the noncondensate and of the degenerate Fermi
gas induced by the condensate oscillations. The back action of these
fluctuations on the condensate motion is then calculated perturbatively to
second order in the interaction coupling constant to obtain the explicit
expression for frequency shifts and damping rates.

Our starting point is the trapped binary Bose-Fermi mixture that is
portrayed as a thermodynamic equilibrium system under the grand canonical
ensemble whose thermodynamic variables are $N_b$ and $N_f$, respectively,
the total number of trapped bosonic and fermionic atoms, $T$, the absolute
temperature, and $\mu _b$ and $\mu _f$, the chemical potentials. In terms of
the creation and annihilation bosonic (fermionic) field operators $\psi ^{+}(%
{\bf r},t)$ and $\psi ({\bf r},t)$ ($\phi ^{+}({\bf r},t)$ and $\phi ({\bf r}%
,t)$), the density Hamiltonian of the system takes the form (in units of $%
\hbar =1$, and all field operators depend on ${\bf r}$ and $t$)
\begin{eqnarray}
{\cal H} &=&{\cal H}_b+{\cal H}_f+{\cal H}_{bf},  \nonumber \\
{\cal H}_b &=&\psi ^{+}\left[ -\frac{{\bf \bigtriangledown }^2}{2m_b}%
+V_{trap}^b({\bf r})-\mu _b\right] \psi +\frac{g_{bb}}2\psi ^{+}\psi
^{+}\psi \psi ,  \nonumber \\
{\cal H}_f &=&\phi ^{+}\left[ -\frac{{\bf \bigtriangledown }^2}{2m_f}%
+V_{trap}^f({\bf r})-\mu _f\right] \phi ,  \nonumber \\
{\cal H}_{bf} &=&g_{bf}\psi ^{+}\psi \phi ^{+}\phi .  \label{hami}
\end{eqnarray}
Here we consider a {\it spherically} symmetric system, with static external
potentials $V_{trap}^{b,f}({\bf r})=m_{b,f}\omega _{b,f}^2r^2/2$, where $%
m_{b,f}$ are the atomic masses, and $\omega _{b,f}$ are the trap
frequencies. The interaction between bosons and between bosons and
fermions are described by the contact potentials and are
parameterized by the coupling constants $g_{bb}=4\pi \hbar
^2a_{bb}/m_b$ and $g_{bf}=2\pi \hbar ^2a_{bf}/m_r$, to the lowest
order in the $s$-wave scattering length $a_{bb}$ and $a_{bf}$,
with $m_r=m_bm_f/(m_b+m_f)$ being the reduced mass.

\subsection{time-dependent mean-field approximation}

According to the usual treatment for Bose system with broken gauge symmetry,
we shall apply the decomposition: $\psi ({\bf r},t)=\Phi ({\bf r},t)+\tilde{%
\psi}({\bf r},t),$ where $\Phi ({\bf r},t)\equiv \left\langle \psi ({\bf r}%
,t)\right\rangle $ represents a time-dependent condensate wave
function and allows us to describe situations in which the system
is displaced from equilibrium and the condensate is oscillating in
time. With this respect the average $\left\langle ...\right\rangle
$ is intended to be a non-equilibrium average, while
time-independent equilibrium averages will be indicated in this
paper with the symbol $\left\langle ...\right\rangle _0$. The
field operator $\tilde{\psi}({\bf r},t)$ plays the role of
excitations out of the
condensate, and by definition satisfies the condition $\left\langle \tilde{%
\psi}({\bf r},t)\right\rangle =0$. This ansatz is then inserted in the
equation of motion for $\psi ({\bf r},t)$:
\begin{eqnarray}
i\frac \partial {\partial t}\psi &=&\left[ -\frac{{\bf \bigtriangledown }^2}{%
2m_b}+V_{trap}^b({\bf r})-\mu _b\right] \psi  \nonumber \\
&&+g_{bb}\psi ^{+}\psi \psi +g_{bf}\psi \phi ^{+}\phi .  \label{eom}
\end{eqnarray}
Taking a statistical average over Eq. (\ref{eom}) and setting the triplet
average values $\left\langle \tilde{\psi}^{+}({\bf r},t)\tilde{\psi}({\bf r}%
,t)\tilde{\psi}({\bf r},t)\right\rangle $ and $\left\langle \tilde{\psi}(%
{\bf r},t)\phi ^{+}({\bf r},t)\phi ({\bf r},t)\right\rangle $ to zero \cite
{note} thus leads to the following equation of motion for the condensate
wave function
\begin{eqnarray}
i\frac \partial {\partial t}\Phi ({\bf r},t) &=&\left[ -\frac{{\bf %
\bigtriangledown }^2}{2m_b}+V_{trap}^b({\bf r})-\mu _b\right] \Phi ({\bf r}%
,t)  \nonumber \\
&&+g_{bb}\left| \Phi ({\bf r},t)\right| ^2\Phi ({\bf r},t)+2g_{bb}\tilde{n}(%
{\bf r},t)\Phi ({\bf r},t)  \nonumber \\
&&+g_{bb}\tilde{m}({\bf r},t)\Phi ^{*}({\bf r},t)+g_{bf}n_f({\bf r},t)\Phi (%
{\bf r},t),  \label{TDGPE}
\end{eqnarray}
where the densities are defined, respectively, as $\tilde{n}({\bf r}%
,t)\equiv \left\langle \tilde{\psi}^{+}({\bf r},t)\tilde{\psi}({\bf r}%
,t)\right\rangle $, $\tilde{m}({\bf r},t)\equiv \left\langle \tilde{\psi}(%
{\bf r},t)\tilde{\psi}({\bf r},t)\right\rangle ,$ and $n_f({\bf r},t)\equiv
\left\langle \phi ^{+}({\bf r},t)\phi ({\bf r},t)\right\rangle $. Under the
stationary condition, we replace $\Phi ({\bf r},t)$, $\tilde{n}({\bf r},t)$,
$\tilde{m}({\bf r},t)$ and $n_f({\bf r},t)$ by their equilibrium values $%
\Phi _0({\bf r})\equiv \left\langle \psi ({\bf r})\right\rangle _0$, $\tilde{%
n}^0({\bf r})\equiv \left\langle \tilde{\psi}^{+}({\bf r})\tilde{\psi}({\bf r%
})\right\rangle _0$, $\tilde{m}^0({\bf r})\equiv \left\langle \tilde{\psi}(%
{\bf r})\tilde{\psi}({\bf r})\right\rangle _0$ and $n_f^0({\bf r})\equiv
\left\langle \phi ^{+}({\bf r})\phi ({\bf r})\right\rangle _0$,
respectively. This yields the generalized time-independent Gross-Pitaevskii
(GP) equation for Bose-Fermi mixtures \cite{hu}
\begin{equation}
\left[ -\frac{{\bf \bigtriangledown }^2}{2m_b}+V_{trap}^b-\mu
_b+g_{bb}\left( n_0+2\tilde{n}^0\right) +g_{bf}n_f^0\right] \Phi _0=0,
\label{GPE}
\end{equation}
where $n_0({\bf r})=\left| \Phi _0({\bf r})\right| ^2$ is the condensate
density. In the above equation, we already use the Popov prescription: $%
\tilde{m}^0({\bf r})=0,$ which amounts to neglect the effects arising from
the equilibrium anomalous density \cite{note2}.

We are interested in the small amplitude oscillations of the condensate,
which is only slightly displaced from its stationary value $\Phi _0({\bf r})$%
: $\Phi ({\bf r},t)=\Phi _0({\bf r})+\delta \Phi ({\bf r},t),$ where $\delta
\Phi ({\bf r},t)$ is a small fluctuation. This small oscillations can
consequently induce small fluctuations of the densities around their
equilibrium values: $\tilde{n}({\bf r},t)=\tilde{n}^0({\bf r})+\delta \tilde{%
n}({\bf r},t)$, $\tilde{m}({\bf r},t)=\delta \tilde{m}({\bf r},t)$, and $n_f(%
{\bf r},t)=n_f^0({\bf r})+\delta n_f({\bf r},t)$. The time-dependent
equation for $\delta \Phi ({\bf r},t)$ is then obtained by linearizing the
equation of motion (\ref{TDGPE})
\begin{eqnarray}
i\frac \partial {\partial t}\delta \Phi ({\bf r},t) &=&{\cal L}\delta \Phi (%
{\bf r},t)+g_{bb}n_0({\bf r})\delta \Phi ^{*}({\bf r},t)  \nonumber \\
&&+2g_{bb}\Phi _0({\bf r})\delta \tilde{n}({\bf r},t)+g_{bb}\Phi _0({\bf r}%
)\delta \tilde{m}({\bf r},t)  \nonumber \\
&&+g_{bf}\Phi _0({\bf r})\delta n_f({\bf r},t),  \label{eomdfai}
\end{eqnarray}
where we have introduced the Hermitian operator
\begin{equation}
{\cal L\equiv }-\frac{{\bf \bigtriangledown }^2}{2m_b}+V_{trap}^b\left( {\bf %
r}\right) -\mu _b+2g_{bb}n_b^0({\bf r})+g_{bf}n_f^0({\bf r}),  \label{lgp}
\end{equation}
and $n_b^0({\bf r})=n_0({\bf r})+\tilde{n}^0({\bf r})$, the total density of
bosons. In Eq. (\ref{eomdfai}), the terms containing $\delta \tilde{n}$, $%
\delta \tilde{m}$ and $\delta n_f$ account for the {\em dynamic coupling}
between the condensate and the fluctuations of the noncondensate component
and of the degenerate Fermi gas. Assuming that the condensate oscillates
with frequency $\omega $: $\delta \Phi ({\bf r},t)=\delta \Phi ({\bf r}%
)e^{-i\omega t}$ and $\delta \Phi ^{*}({\bf r},t)=\delta \Phi ^{*}({\bf r}%
)e^{-i\omega t}$ (note that $\delta \Phi ({\bf r})$ and $\delta \Phi ^{*}(%
{\bf r})$ are independent), and consequently $\delta \tilde{n}({\bf r}%
,t)=\delta \tilde{n}({\bf r})e^{-i\omega t}$, $\delta \tilde{m}({\bf r},t)=$
$\delta \tilde{m}({\bf r})e^{-i\omega t}$ and $\delta n_f({\bf r},t)=$ $%
\delta n_f({\bf r})e^{-i\omega t}$, one finds
\begin{eqnarray}
\omega \delta \Phi ({\bf r}) &=&{\cal L}\delta \Phi ({\bf r})+g_{bb}n_0({\bf %
r})\delta \Phi ^{*}({\bf r})  \nonumber \\
&&+2g_{bb}\Phi _0({\bf r})\delta \tilde{n}({\bf r})+g_{bb}\Phi _0({\bf r}%
)\delta \tilde{m}({\bf r})  \nonumber \\
&&+g_{bf}\Phi _0({\bf r})\delta n_f({\bf r}).  \label{eomdfai2}
\end{eqnarray}
In the absence of coupling terms, Eq. (\ref{eomdfai2}) and its adjoint are
formally equivalent to the time-independent Bogoliubov-deGennes (BdG)
equations \cite{hu}
\begin{equation}
\omega _i^B\left(
\begin{array}{c}
u_i({\bf r}) \\
-v_i({\bf r})
\end{array}
\right) =\left(
\begin{array}{cc}
{\cal L} & g_{bb}n_0({\bf r}) \\
g_{bb}n_0({\bf r}) & {\cal L}
\end{array}
\right) \left(
\begin{array}{c}
u_i({\bf r}) \\
v_i({\bf r})
\end{array}
\right) ,  \label{BdG}
\end{equation}
which define the Bogoliubov quasiparticle wave functions $u_i({\bf r})$ and $%
v_i({\bf r})$ with excitation energies $\omega _i^B$. This equivalence is
not surprising since the Bose broken symmetry leads quite generally to the
one-one correspondence between the small oscillations of the condensate and
the single-quasiparticle wave functions \cite{bks}. For the purpose of
solving Eq. (\ref{eomdfai2}), to leading order of the two coupling constants
\cite{note3}, we thus can select the Bogoliubov quasiparticle wave functions
corresponding to the low-energy collective mode that we are interested, and
set accordingly $\delta \Phi _0({\bf r})=u({\bf r})$, $\delta \Phi _0^{*}(%
{\bf r})=v({\bf r})$ and $\omega =\omega _0$.

The first-order correction due to the fluctuations $\delta \tilde{n}({\bf r}%
) $, $\delta \tilde{m}({\bf r})$ (and its complex conjugate), and $\delta
n_f({\bf r})$, can be calculated by expanding
\begin{eqnarray}
\left(
\begin{array}{c}
\delta \Phi ({\bf r}) \\
\delta \Phi ^{*}({\bf r})
\end{array}
\right) &=&\left(
\begin{array}{c}
u({\bf r}) \\
v({\bf r})
\end{array}
\right) +\left(
\begin{array}{c}
\delta \Phi _1({\bf r}) \\
\delta \Phi _1^{*}({\bf r})
\end{array}
\right) ,  \label{dfai} \\
&&  \nonumber \\
\omega &=&\omega _0+\delta \omega -i\gamma ,  \label{dw}
\end{eqnarray}
where $\delta \omega $ represents the shift in the real part of the
frequency and $\gamma $ is the damping rate. The correction of the wave
functions in Eq. (\ref{dfai}) is chosen to be orthogonal to the unperturbed
Bogoliubov quasiparticle wave functions,
\begin{equation}
\int d{\bf r}\left( u^{*}\left( {\bf r}\right) \delta \Phi _1({\bf r})-v^{*}(%
{\bf r})\delta \Phi _1^{*}({\bf r})\right) =0.  \label{orth}
\end{equation}
Inserting this perturbation ansatz into Eq. (\ref{eomdfai2}) and its
adjoint, we multiply the first equation by $u^{*}\left( {\bf r}\right) $ and
the latter by $v^{*}\left( {\bf r}\right) $, and integrate over space. By
using Eq. (\ref{orth}) and the normalization condition
\begin{equation}
\int d{\bf r}\left( u^{*}\left( {\bf r}\right) u({\bf r})-v^{*}({\bf r})v(%
{\bf r})\right) =1,  \label{norm}
\end{equation}
we get the following relation for the eigenfrequency correction:
\begin{eqnarray}
\delta \omega -i\gamma &=&\int d{\bf r}\Phi _0({\bf r})\left[ 2g_{bb}\left(
u^{*}({\bf r})+v^{*}({\bf r})\right) \delta \tilde{n}({\bf r})\right.
\nonumber \\
&&+g_{bb}\left( u^{*}({\bf r})\delta \tilde{m}({\bf r})+v^{*}({\bf r})\delta
\tilde{m}^{*}({\bf r})\right)  \nonumber \\
&&+\left. g_{bf}\left( u^{*}({\bf r})+v^{*}({\bf r})\right) \delta n_f({\bf r%
})\right] .  \label{c1}
\end{eqnarray}
In the next subsection, based on the RPA and linear response theory, we will
derive the explicit expressions for $\delta \tilde{n}({\bf r})$, $\delta
\tilde{m}({\bf r})$, $\delta \tilde{m}^{*}({\bf r})$ and $\delta n_f({\bf r}%
) $, which are induced by the condensate oscillations. An alternative way to
get these expressions in case of pure Bose gases has been outlined by
Giorgini in Ref. \cite{giorgini}. Our derivation presented below is somewhat
simpler and more transparent in physics.

\subsection{RPA and linear response theory}

Let us consider the interaction terms in the density Hamiltonian (\ref{hami}%
) that couples the condensate wave function to the noncondensate component
and the degenerate Fermi gas
\begin{eqnarray}
{\cal H}_{int} &=&\frac{g_{bb}}2\int d{\bf r}\left[ 4\left| \Phi ({\bf r}%
,t)\right| ^2\tilde{\psi}^{+}({\bf r},t)\tilde{\psi}({\bf r},t)\right.
\nonumber \\
&&\left. +\Phi ^{*2}({\bf r},t)\tilde{\psi}({\bf r},t)\tilde{\psi}({\bf r}%
,t)+\Phi ^2({\bf r},t)\tilde{\psi}^{+}({\bf r},t)\tilde{\psi}^{+}({\bf r}%
,t)\right]  \nonumber \\
&&+g_{bf}\int d{\bf r}\left| \Phi ({\bf r},t)\right| ^2\phi ^{+}({\bf r}%
,t)\phi ({\bf r},t).  \label{inthami}
\end{eqnarray}
Consistent with setting the triplet averages to zero in derivation of Eq. (%
\ref{TDGPE}), we have dropped the terms linear in $\Phi ({\bf r},t)$. In the
spirit of RPA, by linearizing the above interaction Hamiltonian, we identify
the perturbation induced by small amplitude-oscillations of the condensate
(the $e^{-i\omega t}$ dependence is not shown explicitly):
\begin{eqnarray}
{\cal H}_{int}^{perb} &=&g_{bb}\int d{\bf r}\Phi _0({\bf r})\left[ 2\left( u(%
{\bf r})+v({\bf r})\right) \tilde{\psi}^{+}({\bf r},t)\tilde{\psi}({\bf r}%
,t)\right.  \nonumber \\
&&\left. +v({\bf r})\tilde{\psi}({\bf r},t)\tilde{\psi}({\bf r},t)+u({\bf r})%
\tilde{\psi}^{+}({\bf r},t)\tilde{\psi}^{+}({\bf r},t)\right]  \nonumber \\
&&+g_{bf}\int d{\bf r}\Phi _0({\bf r})\left( u({\bf r})+v({\bf r})\right)
\phi ^{+}({\bf r},t)\phi ({\bf r},t),  \label{perbhami}
\end{eqnarray}
where to the leading order we have replaced $\delta \Phi ({\bf r},t)$ and $%
\delta \Phi ^{*}({\bf r},t)$, respectively, by $u({\bf r})e^{-i\omega t}$
and $v({\bf r})e^{-i\omega t}$. Within the linear response theory, the
fluctuations are given by
\begin{equation}
\left(
\begin{array}{c}
\delta \tilde{n} \\
\delta \tilde{m} \\
\delta \tilde{m}^{*}
\end{array}
\right) =g_{bb}\left(
\begin{array}{ccc}
\chi _{\tilde{n}\tilde{n}} & \chi _{\tilde{n}\tilde{m}} & \chi _{\tilde{n}%
\tilde{m}^{+}} \\
\chi _{\tilde{m}\tilde{n}} & \chi _{\tilde{m}\tilde{m}} & \chi _{\tilde{m}%
\tilde{m}^{+}} \\
\chi _{\tilde{m}^{+}\tilde{n}} & \chi _{\tilde{m}^{+}\tilde{m}} & \chi _{%
\tilde{m}^{+}\tilde{m}^{+}}
\end{array}
\right) \left(
\begin{array}{c}
2\Phi _0\left( u+v\right) \\
\Phi _0v \\
\Phi _0u
\end{array}
\right)  \label{dnmm}
\end{equation}
and
\begin{equation}
\delta n_f=g_{bf}\int d{\bf r}^{\prime }\chi _f\left( {\bf r},{\bf r}%
^{\prime };\omega \right) \Phi _0({\bf r}^{\prime })\left( u({\bf r}^{\prime
})+v({\bf r}^{\prime })\right) .  \label{dnf}
\end{equation}
Here we define $\chi _{\alpha \beta }\Phi _0u\equiv \int d{\bf r}^{\prime
}\chi _{\alpha \beta }\left( {\bf r},{\bf r}^{\prime };\omega \right) \Phi
_0({\bf r}^{\prime })u({\bf r}^{\prime })$ and $\chi _{\alpha \beta }\Phi
_0v\equiv \int d{\bf r}^{\prime }\chi _{\alpha \beta }\left( {\bf r},{\bf r}%
^{\prime };\omega \right) \Phi _0({\bf r}^{\prime })v({\bf r}^{\prime })$ in
Eq. (\ref{dnmm}), where the indices $\alpha ,\beta =\tilde{n},\tilde{m},$ or
$\tilde{m}^{+}$. $\chi _{\alpha \beta }$ and $\chi _f$ are the usual
two-particle correlation functions for the Bose and Fermi gas \cite{fetter}.
By using Wick's theorem, they can be easily expressed in terms of the
quasiparticle energies and wave functions. For instance, for $\chi _{\tilde{n%
}\tilde{n}}$, with the help of the Bogoliubov transformation we can write $%
\tilde{\psi}({\bf r},t)=\sum_i\left[ u_i({\bf r})\hat{\alpha}_ie^{-i\omega
_i^Bt}+v_i^{*}({\bf r})\hat{\alpha}_i^{+}e^{i\omega _i^Bt}\right] $ in terms
of the Bogoliubov quasiparticle operators $\hat{\alpha}_i$ and $\hat{\alpha}%
_i^{+}$. It is then straightforward to obtain
\begin{eqnarray}
\chi _{\tilde{n}\tilde{n}} &=&\chi _{\tilde{n}\tilde{n}}^{(1)}\left( {\bf r},%
{\bf r}^{\prime };\omega \right) +\chi _{\tilde{n}\tilde{n}}^{(2)}\left(
{\bf r},{\bf r}^{\prime };\omega \right) ,  \label{kapann} \\
\chi _{\tilde{n}\tilde{n}}^{(1)} &=&\sum\limits_{ij}\frac{\left(
u_i^{*}u_j+v_i^{*}v_j\right) \left( u_iu_j^{*}+v_iv_j^{*}\right) \left(
f_i^B-f_j^B\right) }{\omega ^{+}+\left( \omega _i^B-\omega _j^B\right) },
\nonumber \\
\chi _{\tilde{n}\tilde{n}}^{(2)} &=&\frac 12\sum\limits_{ij}\left[ \frac{%
\left( u_iv_j+v_iu_j\right) \left( u_i^{*}v_j^{*}+v_i^{*}u_j^{*}\right)
\left( 1+f_i^B+f_j^B\right) }{\omega ^{+}-\left( \omega _i^B+\omega
_j^B\right) }\right.  \nonumber \\
&&\left. -\frac{\left( u_i^{*}v_j^{*}+v_i^{*}u_j^{*}\right) \left(
u_iv_j+v_iu_j\right) \left( 1+f_i^B+f_j^B\right) }{\omega ^{+}+\left( \omega
_i^B+\omega _j^B\right) }\right] ,  \nonumber
\end{eqnarray}
where $\omega ^{+}=\omega +i0^{+}$. We have used the abbreviation: $%
u_i^{*}u_ju_iu_j^{*}=u_i^{*}({\bf r})u_j({\bf r})u_i({\bf r}^{\prime
})u_j^{*}({\bf r}^{\prime })$, {\it etc.}, and $f_i^B=\left\langle \hat{%
\alpha}_i^{+}\hat{\alpha}_i\right\rangle _0=1/\left( e^{\beta \omega
_i^B}-1\right) $ is the Bose-Einstein distribution function with $\beta
=1/k_BT$. $\chi _{\tilde{n}\tilde{n}}^{(1)}$ and $\chi _{\tilde{n}\tilde{n}%
}^{(2)}$ correspond, respectively, to the excitation of single
quasiparticles and of pairs of quasiparticles. For $\chi _f$, we have
\begin{equation}
\chi _f=\sum\limits_{ij}\left( f_i^F-f_j^F\right) \frac{\varphi _i^{*}({\bf r%
})\varphi _j({\bf r})\varphi _i({\bf r}^{\prime })\varphi _j^{*}({\bf r}%
^{\prime })}{\omega ^{+}+\left( \omega _i^F-\omega _j^F\right) },
\label{kapaf}
\end{equation}
where $f_i^F=1/\left( e^{\beta \omega _i^F}+1\right) $ is the Fermi-Dirac
distribution, and the single-particle wave function $\varphi _i({\bf r})$
satisfies the stationary Schr\"{o}dinger equation \cite{hu}
\begin{equation}
\left[ -\frac{{\bf \bigtriangledown }^2}{2m_f}+V_{trap}^f\left( {\bf r}%
\right) -\mu _f+g_{bf}n_b^0\left( {\bf r}\right) \right] \varphi _i=\omega
_i^F\varphi _i.  \label{dfg}
\end{equation}

\subsection{eigenfrequency correction}

Substituting the fluctuations (\ref{dnmm}) and (\ref{dnf}) into Eq. (\ref{c1}%
), and using the explicit expressions for $\chi _{\alpha \beta }$ and $\chi
_f$, one finds to second order of $g_{bb}$ and $g_{bf}$,
\begin{eqnarray}
\delta \omega -i\gamma &=&4g_{bb}^2\sum\limits_{ij}\left( f_i^B-f_j^B\right)
\frac{\left| A_{ij}\right| ^2}{\omega ^{+}+\left( \omega _i^B-\omega
_j^B\right) }  \nonumber \\
&&+2g_{bb}^2\sum\limits_{ij}\left( 1+f_i^B+f_j^B\right) \left( \frac{\left|
B_{ij}\right| ^2}{\omega ^{+}-\left( \omega _i^B+\omega _j^B\right) }\right.
\nonumber \\
&&\left. -\frac{\left| \tilde{B}_{ij}\right| ^2}{\omega ^{+}+\left( \omega
_i^B+\omega _j^B\right) }\right)  \nonumber \\
&&+g_{bf}^2\sum\limits_{ij}\left( f_i^F-f_j^F\right) \frac{\left|
C_{ij}\right| ^2}{\omega ^{+}+\left( \omega _i^F-\omega _j^F\right) },
\label{c2}
\end{eqnarray}
where the matrix elements $A_{ij}$, $B_{ij}$, $\tilde{B}_{ij}$, and $C_{ij}$
are, respectively, given by
\begin{eqnarray}
A_{ij} &=&\int d{\bf r}\Phi _0\left[ u\left(
u_iu_j^{*}+v_iv_j^{*}+v_iu_j^{*}\right) \right.  \nonumber \\
&&\left. +v\left( u_iu_j^{*}+v_iv_j^{*}+u_iv_j^{*}\right) \right] ,
\nonumber \\
B_{ij} &=&\int d{\bf r}\Phi _0\left[ u\left(
u_i^{*}v_j^{*}+v_i^{*}u_j^{*}+u_i^{*}u_j^{*}\right) \right.  \nonumber \\
&&\left. +v\left( u_i^{*}v_j^{*}+v_i^{*}u_j^{*}+v_i^{*}v_j^{*}\right)
\right] ,  \nonumber \\
\tilde{B}_{ij} &=&\int d{\bf r}\Phi _0\left[ u\left(
u_iv_j+v_iu_j+v_iv_j\right) \right.  \nonumber \\
&&\left. +v\left( u_iv_j+v_iu_j+u_iu_j\right) \right] ,  \nonumber \\
C_{ij} &=&\int d{\bf r}\Phi _0({\bf r})\left( u\left( {\bf r}\right)
+v\left( {\bf r}\right) \right) \varphi _i\left( {\bf r}\right) \varphi
_j^{*}\left( {\bf r}\right) .  \label{abbc}
\end{eqnarray}
Eqs. (\ref{c2}) and (\ref{abbc}) are the main result of this section.
Without the fermionic component, these equations coincides with the finding
obtained by Giorgini (the Eqs. (39) and (40) in the second paper of Ref.
\cite{giorgini} ) as they should be. The last term in the right-hand side of
Eq. (\ref{c2}) is novel and arises from the many possibilities of
independent particle-hole excitations \cite{note4}. This mechanism is known
as Landau damping due to the Bose-Fermi coupling. On the other hand, the
first and second terms in the right-hand side of Eq. (\ref{c2}) correspond,
respectively, to the Landau and Beliaev processes due to the interaction
between bosons \cite{giorgini}.

One of the advantages of our derivation presented here is that to obtain the
eigenfrequency correction we don't need to impose any constraint used in
solving the equilibrium problem, {\it i.e.}, the Popov prescription $\tilde{m%
}^0({\bf r})=0$. For a pure Bose gas at high temperatures close to $T_c$, it
might be reasonable to use the Hartree-Fock spectrum for $\chi _{\alpha
\beta }$. As a result, only $\chi _{\tilde{n}\tilde{n}}$ is nonzero and the
eigenfrequency correction reads
\begin{eqnarray}
\delta \omega -i\gamma &=&4g_{bb}^2\int d{\bf r}\int d{\bf r}^{\prime }\Phi
_0\left( {\bf r}\right) \left( u^{*}\left( {\bf r}\right) +v^{*}\left( {\bf r%
}\right) \right)  \nonumber \\
&&\times \chi _{\tilde{n}\tilde{n}}\left( {\bf r},{\bf r}^{\prime };\omega
\right) \Phi _0\left( {\bf r}^{\prime }\right) \left( u\left( {\bf r}%
^{\prime }\right) +v\left( {\bf r}^{\prime }\right) \right) ,  \label{c3}
\end{eqnarray}
which is identical to the finding of Reidl {\it et al}. obtained by using
the dielectric formalism (the Eq. (52) in Ref. \cite{reidl}), if one notices
that $\delta n_c\left( {\bf r}\right) =\Phi _0\left( {\bf r}\right) \left(
u\left( {\bf r}\right) +v\left( {\bf r}\right) \right) $.

It should be noted that the second term in the right-hand side of Eq. (\ref
{c2}), corresponding to the Beliaev process, is ultraviolet divergent. This
reflects the fact that the contact interaction is an effective low-energy
interaction invalid for high energies. One way to remove this divergence is
to express the coupling constant $g_{bb}$ in terms of the two-body
scattering matrix obtained from the Lippman-Schwinger equation. This
renormalization scheme has been put forward in Ref. \cite{giorgini},
however, only valid in the thermodynamic limit, where the Thomas-Fermi
approximation can be implemented \cite{giorgini}. In our derivation, one can
explicitly show that such divergence comes from the two correlation
functions: $\chi _{\tilde{m}\tilde{m}^{+}}\left( {\bf r},{\bf r}^{\prime
};\omega \right) $ and $\chi _{\tilde{m}^{+}\tilde{m}}\left( {\bf r},{\bf r}%
^{\prime };\omega \right) $. One thus may wish to remove the divergence by
regularizing $\chi _{\tilde{m}\tilde{m}^{+}}$ and $\chi _{\tilde{m}^{+}%
\tilde{m}}$ in real space in a way similar to that described in Refs. \cite
{bruun,note5}. In this paper, for simplicity we shall {\em neglect} the
second term in the right-hand side of Eq. (\ref{c2}), since it is always
very small compared with other two terms at all temperatures. This treatment
is well justified by the excellent agreement between the experimental result
\cite{jin} and the theoretical prediction by Reidl {\it et al}. \cite{reidl}
for a pure Bose gas, where in the theoretical calculations the Beliaev
process is completely ignored.

There is one last technical issue to resolve: concerning the damping rate,
the terms in Eq. (\ref{c2}) involve a sum over many $\delta $ functions in
energy, which, if interpreted exactly, will tend to be null for discrete
quasiparticle states. Here we shall adopt the strategy of Ref. \cite{das}
and use an expression with a Lorentz profile factor in place of the energy $%
\delta $ function
\begin{eqnarray}
\delta \omega -i\gamma &=&4g_{bb}^2\sum\limits_{ij}\frac{\left(
f_i^B-f_j^B\right) \left| A_{ij}\right| ^2}{\omega _0+\delta \omega +\left(
\omega _i^B-\omega _j^B\right) +i\gamma }  \nonumber \\
&&+g_{bf}^2\sum\limits_{ij}\frac{\left( f_i^F-f_j^F\right) \left|
C_{ij}\right| ^2}{\omega _0+\delta \omega +\left( \omega _i^F-\omega
_j^F\right) +i\gamma },  \label{c4}
\end{eqnarray}
which can be solved iteratively for $\delta \omega $ and $\gamma $. This
expression can be formally obtained from Eq. (\ref{c2}) by assuming that the
perturbed resonance frequency $\omega $ is distributed over a range of
values characterized by a Lorentz profile with a width $\gamma $.

The structure of our calculation is then as follows: First we solve the
unperturbed equilibrium problem for $\omega _0$, $u({\bf r})$ and $v({\bf r}%
) $. This step requires solving a closed set of Eqs. (\ref{GPE}), (\ref{BdG}%
), and (\ref{dfg}), which we have referred to as the ``HFB-Popov'' equations
for dilute Bose-Fermi mixtures. We already have reported on our
self-consistent algorithm in Ref. \cite{hu} for this problem. As a result,
we have all the necessary inputs, namely, $\Phi _0\left( {\bf r}\right) $, $%
\omega _i^B$, $\omega _i^F$, $u_i({\bf r})$ and $v_i({\bf r})$, and $\varphi
_i({\bf r})$ for performing the second step: the use of Eq. (\ref{c4}) in
connection with the matrix elements $A_{ij}$ and $C_{ij}$ defined in Eq. (%
\ref{abbc}).

\section{numerical results}

In this work we analyze the low-lying condensate oscillations of a
Bose-Fermi mixture for varying Bose-Fermi coupling constant and temperature
in an isotropic harmonic trap, for which the order parameter $\Phi _0({\bf r}%
)$, the Bogoliubov quasiparticle amplitudes $u_i({\bf r})$ and $v_i({\bf r})$%
, and the orbits $\varphi _i({\bf r})$ can be classified according
to the number of nodes in the radial solution $n$, the orbital
angular momentum $l$, and its projection $m$. For this sake, we
apply the theory developed in the
proceeding section to mixtures of $^{41}{\rm K}-^{40}${\rm K} and $^{87}{\rm %
Rb}-^{40}${\rm K}. For the former system, we discuss some generic properties
of Bose-Fermi mixtures. In {\em this} case the rigid oscillation of the
center of mass, or the dipole mode, will also be an eigenstate of the
many-body system. As a result, the oscillation frequency will be fixed at
the bare trapping frequency, regardless of any interactions. The fulfillment
of this property, which is usually referred to as the generalized Kohn
theorem, thus provides us a stringent test on the correctness of our
results. The second choice of $^{87}{\rm Rb}-^{40}${\rm K} mixture
corresponds to a specific example available in present experiments \cite
{lens}. In this case we include explicitly the mass difference and the
different oscillator frequencies of the trapping potentials for the two
species, and build the possible experimental relevance of our results.

\subsection{$^{41}{\rm K}-^{40}${\rm K}}

We first consider a mixture of $2\times 10^4$ $^{41}{\rm K}$ (boson) and $%
2\times 10^4$ $^{40}${\rm K} (fermion) atoms with the following set of
parameters: $m_b=m_f=0.649\times 10^{-25}$ {\rm kg}, $\omega _b=\omega
_f=2\pi \times 100$ {\rm Hz}, $a_{bb}=286a_0=15.13$ {\rm nm} \cite{cote},
where $a_0=0.529$ ${\rm \AA }$ is the Bohr radius. We also express the
lengths and energies in terms of the characteristic oscillator length $%
a_{ho}^b=(\hbar /m_b\omega _b)^{1/2}$ and characteristic trap energy $\hbar
\omega _b$, respectively.

\begin{figure}[tbp]
\centerline{\includegraphics[width=5.0cm,angle=-90,clip=]{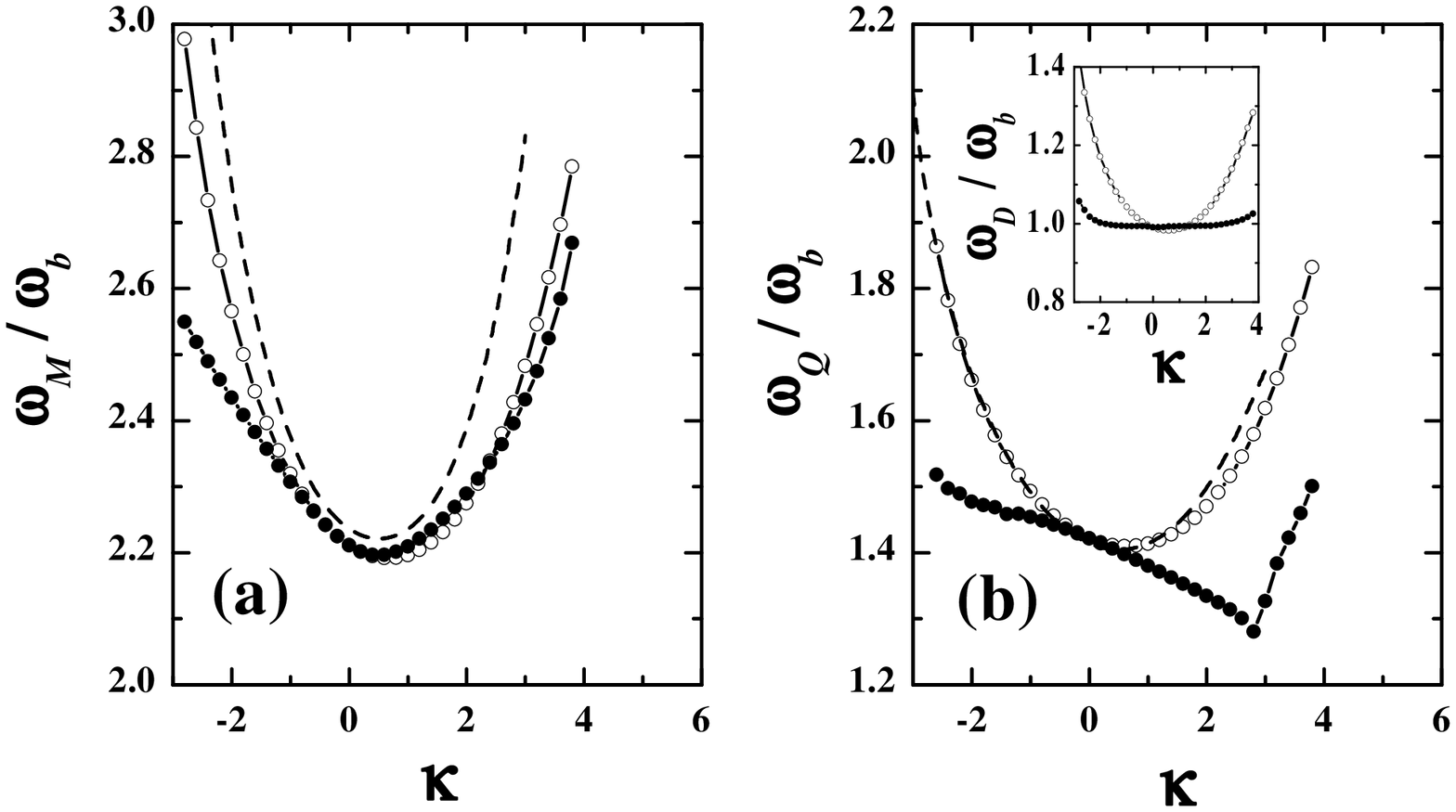}}
\caption{The dispersion relation of the monopole ($l=0$) and quadrupole ($%
l=2 $) condensate oscillations for a mixture composed of $2\times 10^4$ $%
^{41}{\rm K}$ (boson) and $2\times 10^4$ $^{40}${\rm K} (fermion) atoms at $%
T=0.01T_c^0$, where $T_c^0=0.94\omega _bN_b^{1/3}/k_B\approx 122$
{\rm nK} is the critical temperature for an ideal Bose gas in the
thermodynamic limit. The mode frequencies, in units of the bare
trapping frequency, are plotted as a function of the reduced
Bose-Fermi coupling, $\kappa =g_{bf}/g_{bb}$. The lines with open
circles show the unperturbed frequencies calculated by the static
HFB-Popov equations $\omega _0$, while the lines with solid
circles denote $\omega =\omega _0+\delta \omega $. For comparison
the predictions of the scaling theory in Ref. [8] are also plotted
by the dashed lines. The inset in (b) shows the dispersion
relation of the dipole excitation. The other parameters used in
the numerical calculation are: $m_b=m_f=0.649\times 10^{-25}$ {\rm
kg}, $\omega _b=\omega
_f=2\pi \times 100$ {\rm Hz}, and $a_{bb}=286a_0=15.13$ {\rm nm}, where $%
a_0=0.529$ ${\rm \AA }$ is the Bohr radius.} \label{fig1}
\end{figure}

In Figs. (1a) and (1b), we present, respectively, our results for the
monopole ($l=0$) and quadrupole ($l=2$) condensate oscillations at a very
low temperature $T=0.01T_c^0$, where $T_c^0=0.94\omega
_bN_b^{1/3}/k_B\approx 122$ {\rm nK} is the critical temperature for an
ideal Bose gas in the thermodynamic limit. The mode frequencies, in units of
the bare trapping frequency, are plotted as a function of the Bose-Fermi
coupling constant measured relative to the Bose-Bose coupling constant, $%
\kappa =g_{bf}/g_{bb}$. The lines with open circles show the
unperturbed frequencies obtained from the HFB-Popov equations,
$\omega _0$, while the lines with solid circles denote the values
after correction, $\omega =\omega _0+\delta \omega $. For
comparison, the predictions of the scaling theory at zero
temperature are also plotted by the dashed lines \cite{liu}. At
this low temperature, the eigenfrequency shift $\delta \omega $
arises mainly from the dynamics of the degenerate Fermi gas. For
small values of $\left| \kappa \right| \lesssim 1$, $\delta \omega
$ is negligibly small due to square dependence on the Bose-Fermi
coupling constant. However, as $\left| \kappa \right| $ increases
$\delta \omega $ becomes remarkable. In particular, the corrected
frequency for the quadrupole mode decreases with increasing
Bose-Fermi coupling constant up to $\kappa \approx 3$, at which a
sharp upturn occurs. This sharp dip is accompanied by a dramatic
increase of damping rates (not shown in the figure), and can be
well interpreted as a signal to approach the spatial separation
(demixing) point of the two species. On the contrary, the
unperturbed quadrupole frequency $\omega _0$ has a qualitatively
different behavior against $\kappa $: it shows a parabolic
dependence with a minimum located around $\kappa =0.5$. In spite
of its large value at $\left| \kappa \right| \gtrsim 1$, $\delta
\omega $ is still much smaller than unperturbed frequency $\omega
_0$ for all the calculated points. Therefore, the criterium for
the applicability of the perturbation theory in Eqs. (\ref {dfai})
and (\ref{dw}) is justified. In the inset of Fig. (1b) we also
show the result for the dipole mode ($l=1$). As we can see,
although $\omega _0$ deviates significantly from the bare trapping
frequency $\omega _b$ for relatively small values of $\kappa $,
$\omega $ still persists at $\omega _b$
over a wide range of $\kappa $ ({\it i.e.}, $-2\lesssim \kappa \lesssim +3$%
). This is consistent with the generalized Kohn theorem. As a result, the
correctness of our theory and numerical calculations is partly checked.

\begin{figure}[tbp]
\centerline{\includegraphics[width=5.0cm,angle=-90,clip=]{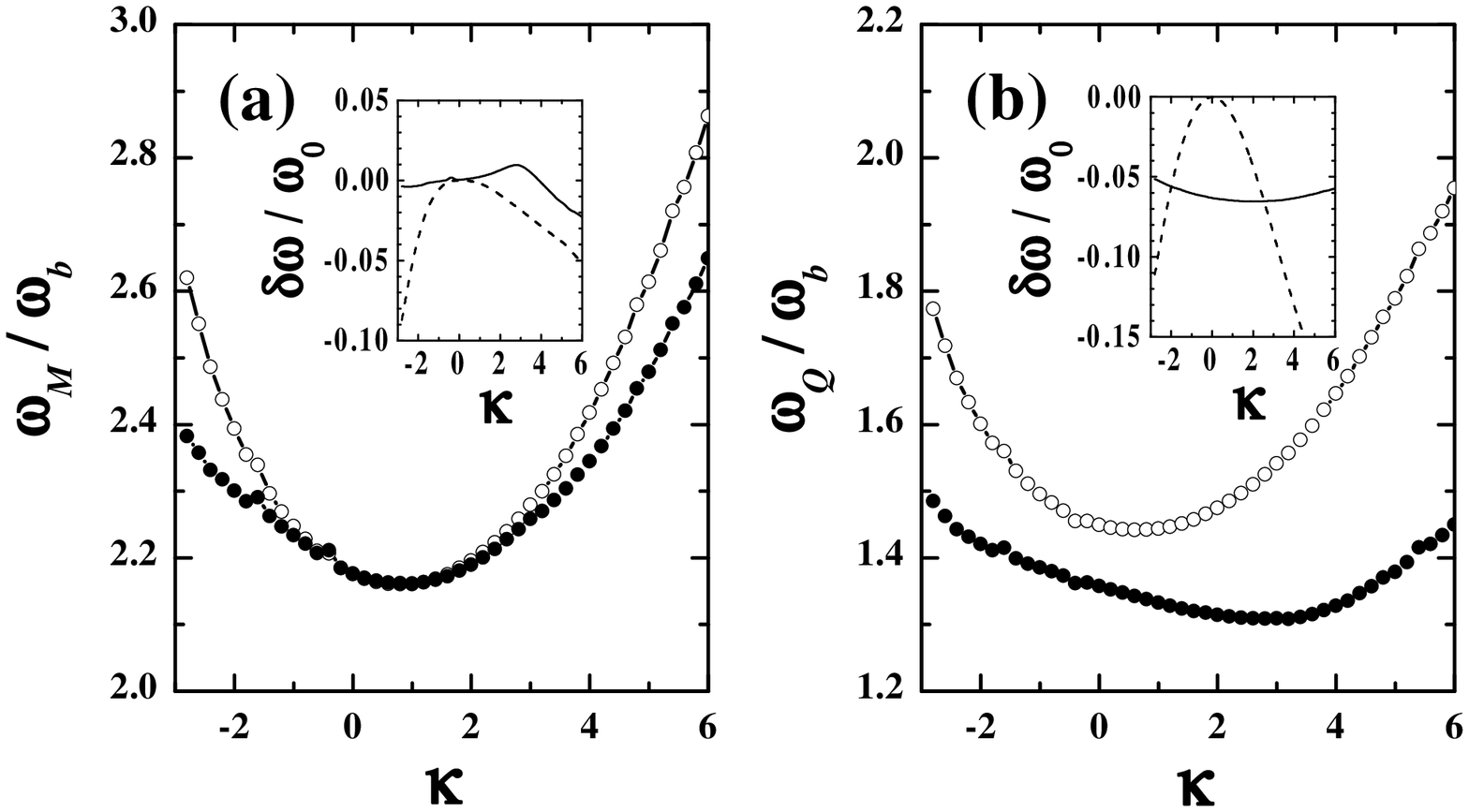}}
\caption{The same as in FIG. 1, but for $T=0.75T_c^0$. In the
insets the solid and dashed lines show, respectively, the
fractional shift $\delta \omega /\omega _0$ due to the first and
second terms in Eq. (\ref{c4}).} \label{fig2}
\end{figure}

The eigenfrequency shifts are affected by the temperature. In Fig.
2, we report the mode frequencies against $\kappa $ at a high temperature $%
T=0.75T_c^0$, where the condensate oscillates in the presence of a large
fraction of above-condensate atoms. Compared with the results for the low
temperature, the eigenfrequency shifts are considerably reduced. The sharp
dip at $\kappa \approx 3$ for the quadrupole mode also becomes much broader.
Moreover, in the absence of the Bose-Fermi coupling, the eigenfrequency
shift is nonzero for the quadrupole mode. This is caused by the dynamics of
the noncondensate component as shown in the inset of Fig. (2b), where the
solid and dashed lines depict, respectively, the fractional shift $\delta
\omega /\omega _0$ due to the Bose-Bose interaction and Bose-Fermi coupling
(or, in other words, due to the first and second terms in Eq. (\ref{c4})).

\begin{figure}[tbp]
\centerline{\includegraphics[width=5.0cm,angle=-90,clip=]{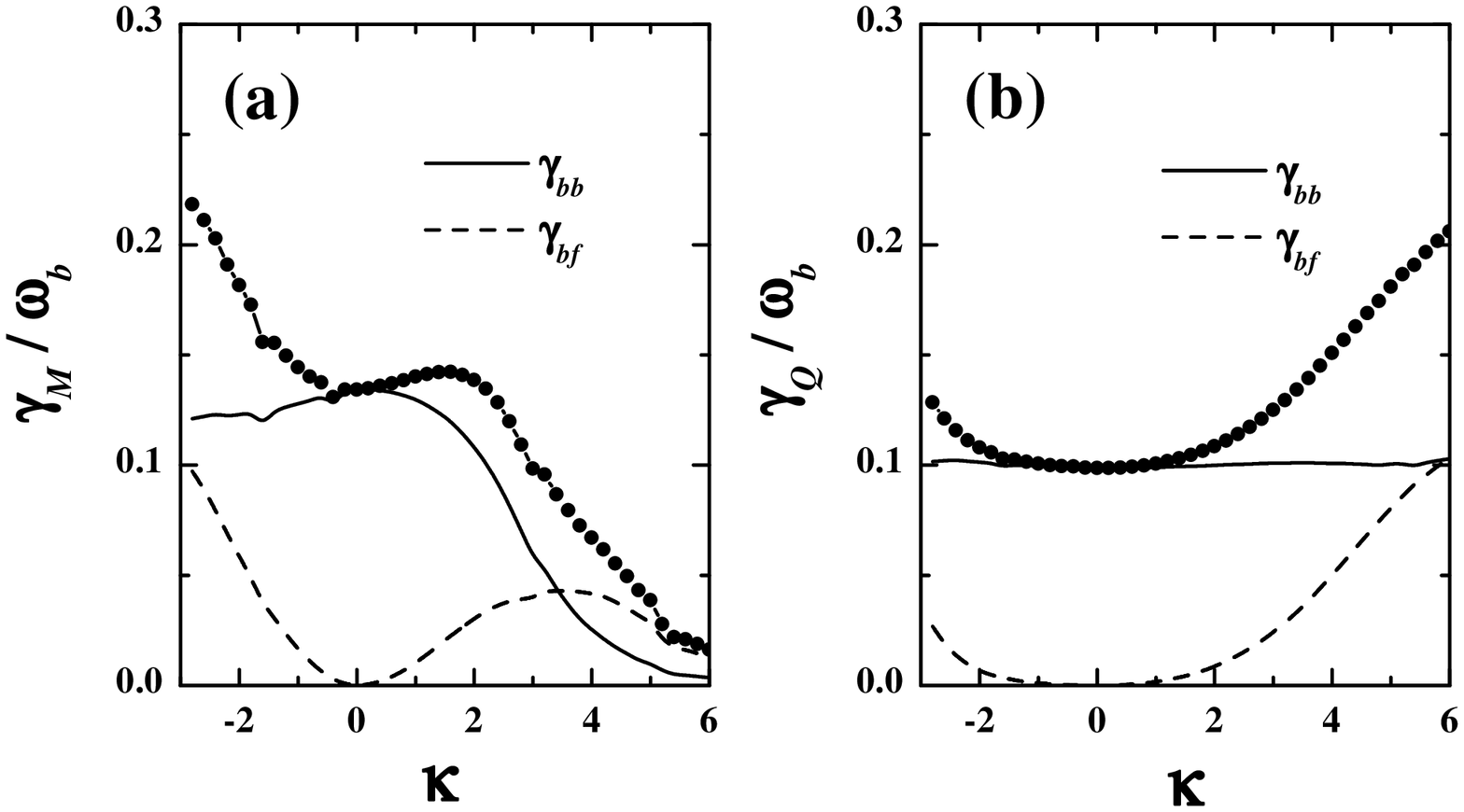}}
\caption{The damping rates of the monopole (a) and quadrupole
condensate oscillations (b) at $T=0.75T_c^0$. The solid and dashed
lines correspond to the contribution from the Landau process due
to the Bose-Bose interaction and due to the Bose-Fermi coupling,
respectively. The line with solid circles is the total
contribution.} \label{fig3}
\end{figure}

\begin{figure}[tbp]
\centerline{\includegraphics[width=5.5cm,angle=-90,clip=]{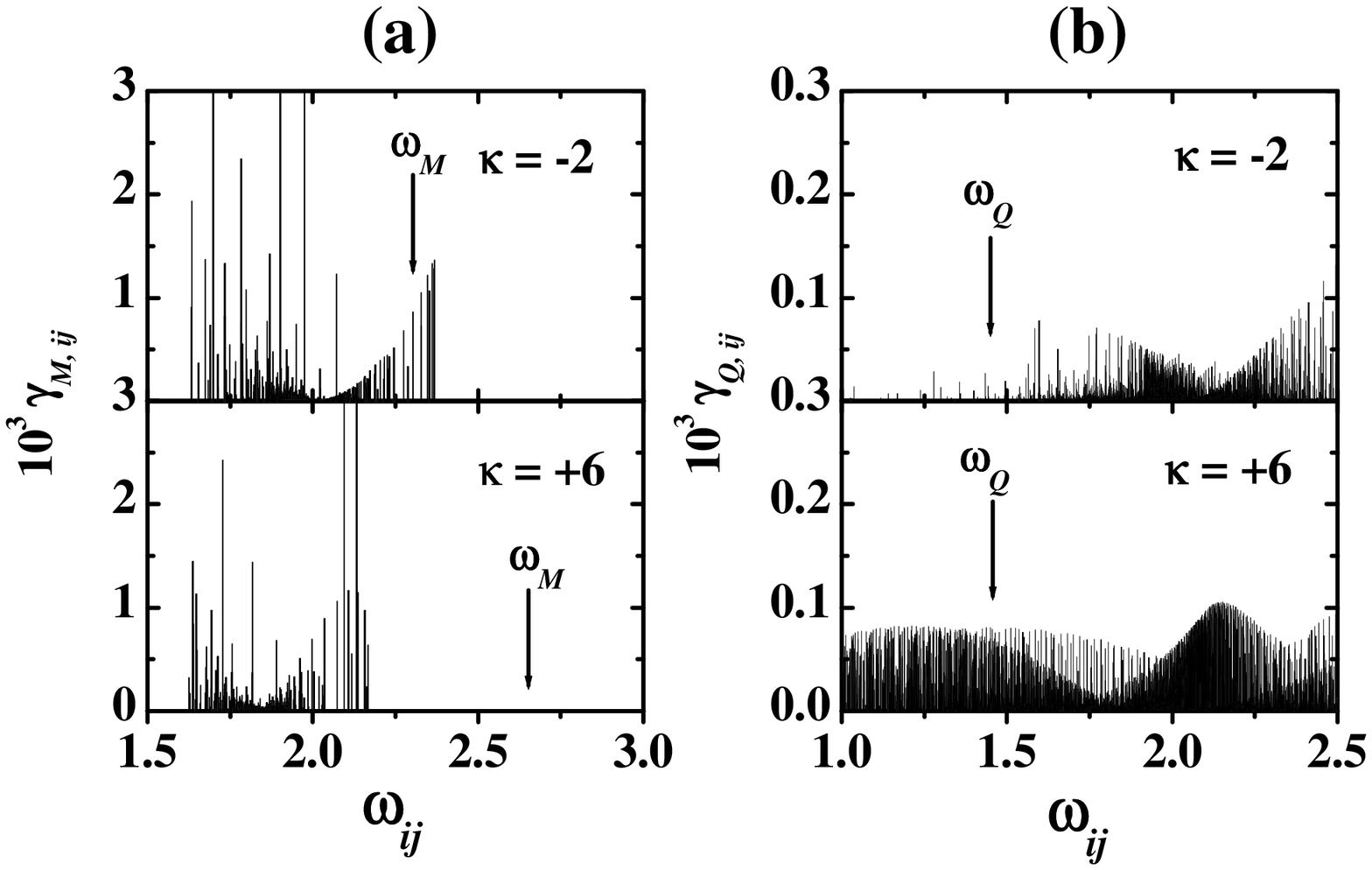}}
\caption{The damping strength $\gamma _{ij\text{ }}$defined in Eq. (\ref{ds}%
) as a function of the transition frequencies $\omega _{ij}$ (in units of $%
\omega _b$) allowed by the selection rules. (a) $\gamma _{ij\text{
}}^{bb}$ for the monopole excitation and (b) $\gamma _{ij\text{
}}^{bf}$ for the quadrupole excitation. In each subplot, upper and
lower panels correspond to the case of $\kappa =-2$ and $\kappa
=+6$, respectively. In the former case the mixture is in mixed
regime, while in the latter the bosonic and fermionic density
profiles separate in space. The arrows point to $\omega =\omega
_0+\delta \omega $.} \label{fig4}
\end{figure}

The damping rate of condensate oscillations at finite temperature
deserves its own study. In Figs. (3a) and (3b), we show,
respectively, our predictions
on the damping rates of monopole and quadrupole oscillations at $T=0.75T_c^0$%
. The lines with solid circles are the sum over two contributions: one is
the Landau damping due to the Bose-Bose interaction (the solid lines), $%
\gamma _{bb}$, and the other is the Landau damping due to the Bose-Fermi
coupling (the dashed line), $\gamma _{bf}$. For the monopole mode, the
essential feature is the decrease of the damping rate as $\kappa $ increases
towards the demixing point. This decrease is mainly attributed by $\gamma
_{bb}$, and reflects the reconstruction of the {\em bosonic}
monopole-excitation spectrum across the demixing point. To better illustrate
this point, we rewrite the expression for $\gamma $ in the following form
\begin{eqnarray}
\gamma &=&\gamma _{bb}+\gamma _{bf},  \label{dr} \\
\gamma _{bb} &=&4g_{bb}^2\omega _0\sum\limits_{ij}\gamma _{ij}^{bb}\frac{%
\gamma /\pi }{\left[ \omega _0+\delta \omega +\left( \omega _i^B-\omega
_j^B\right) \right] ^2+\gamma ^2},  \nonumber \\
\gamma _{bf} &=&g_{bf}^2\omega _0\sum\limits_{ij}\gamma _{ij}^{bf}\frac{%
\gamma /\pi }{\left[ \omega _0+\delta \omega +\left( \omega _i^F-\omega
_j^F\right) \right] ^2+\gamma ^2},  \nonumber
\end{eqnarray}
where the ``damping strength''
\begin{eqnarray}
\gamma _{ij}^{bb} &=&\frac \pi {\omega _0}\left| A_{ij}\right| ^2\left(
f_i^B-f_j^B\right) ,  \nonumber \\
\gamma _{ij}^{bf} &=&\frac \pi {\omega _0}\left| C_{ij}\right| ^2\left(
f_i^F-f_j^F\right) ,  \label{ds}
\end{eqnarray}
have the dimensions of a frequency. In Fig. (4a), we plot $\gamma
_{ij}^{bb}$ against the transition frequency $\omega _{ij}=\omega
_j^B-\omega _i^B>0$ allowed by the selection rules for $\kappa
=-2$ and $\kappa =6$. For the latter value of $\kappa $, the
overlap between the bosonic and fermionic cloud is very small, and
the mixture is deep into the demixing regime. Compared with the
mixing case of $\kappa =-2$, the region of transition frequencies
at $\kappa =6$ narrows, and its center moves to the low-energy
side. Contrarily the calculated $\omega _0+\delta \omega $ has a
blue shift and is completely out of the transition region. As a
consequence, the condensate oscillation is not damped by the
Landau process due to the Bose-Bose interaction. For the
quadrupole mode, instead we observe that the damping rate
increases as the mixture moves towards the demixing point with
increasing $\kappa $. This trend comes from the increase of
$\gamma _{bf}$, and reflects, on the other hand, the
reconstruction of the {\em fermionic} quadrupole-excitation
spectrum. As shown in Fig. (4b), with increasing Bose-Fermi
coupling the damping strength $\gamma _{ij}^{bf}$ becomes larger
and denser. Accordingly, the condensate oscillation is heavily
damped by generating many particle-hole excitations.

\begin{figure}[tbp]
\centerline{\includegraphics[width=5.0cm,angle=-90,clip=]{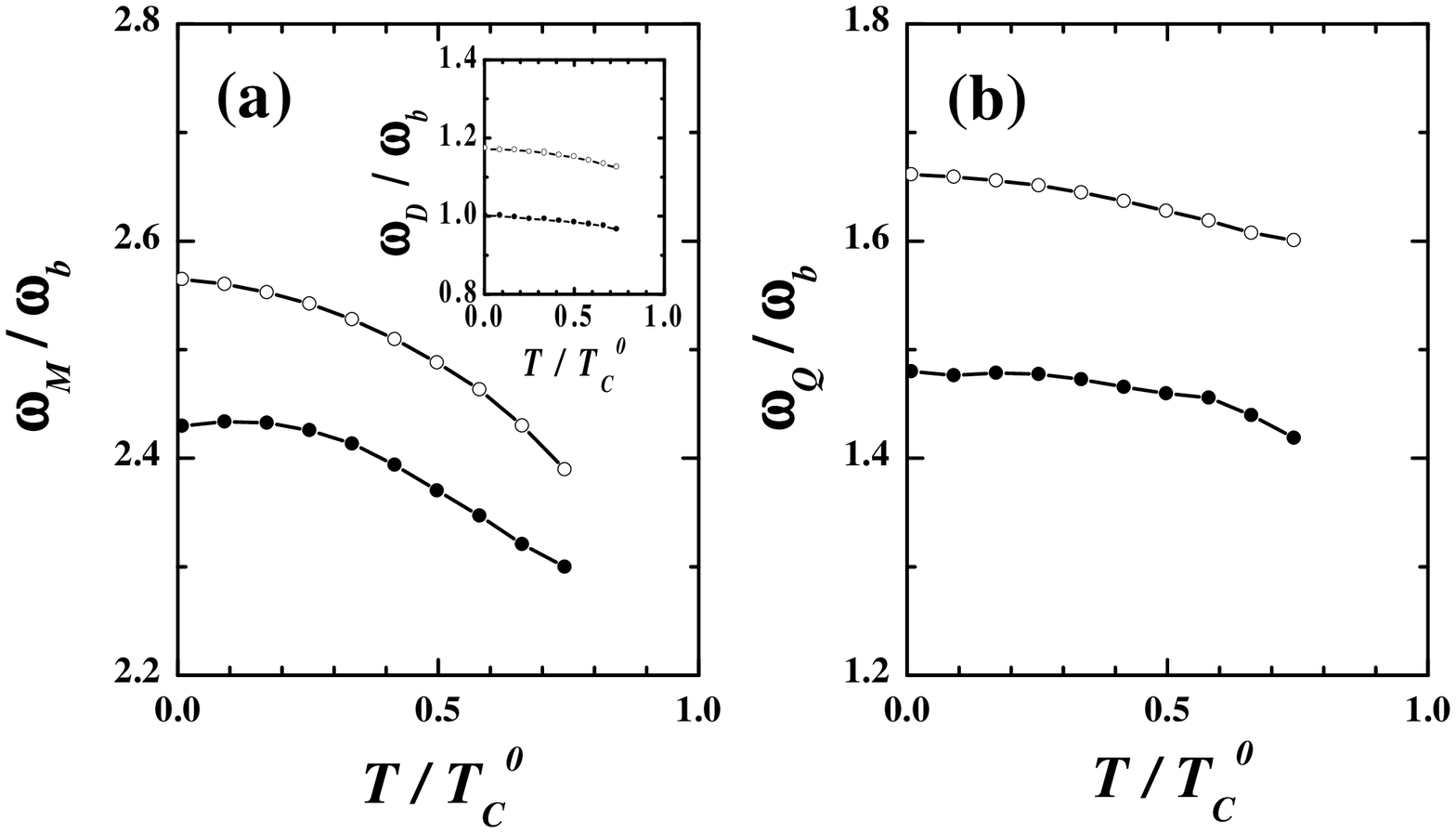}}
\caption{The dispersion relation of the monopole (a) and
quadrupole excitations (b) against the reduce temperature
$T/T_c^0$ at $\kappa = -2$. The inset in (a) shows the dispersion
relation for the dipole mode.} \label{fig5}
\end{figure}

\begin{figure}[tbp]
\centerline{\includegraphics[width=5.0cm,angle=-90,clip=]{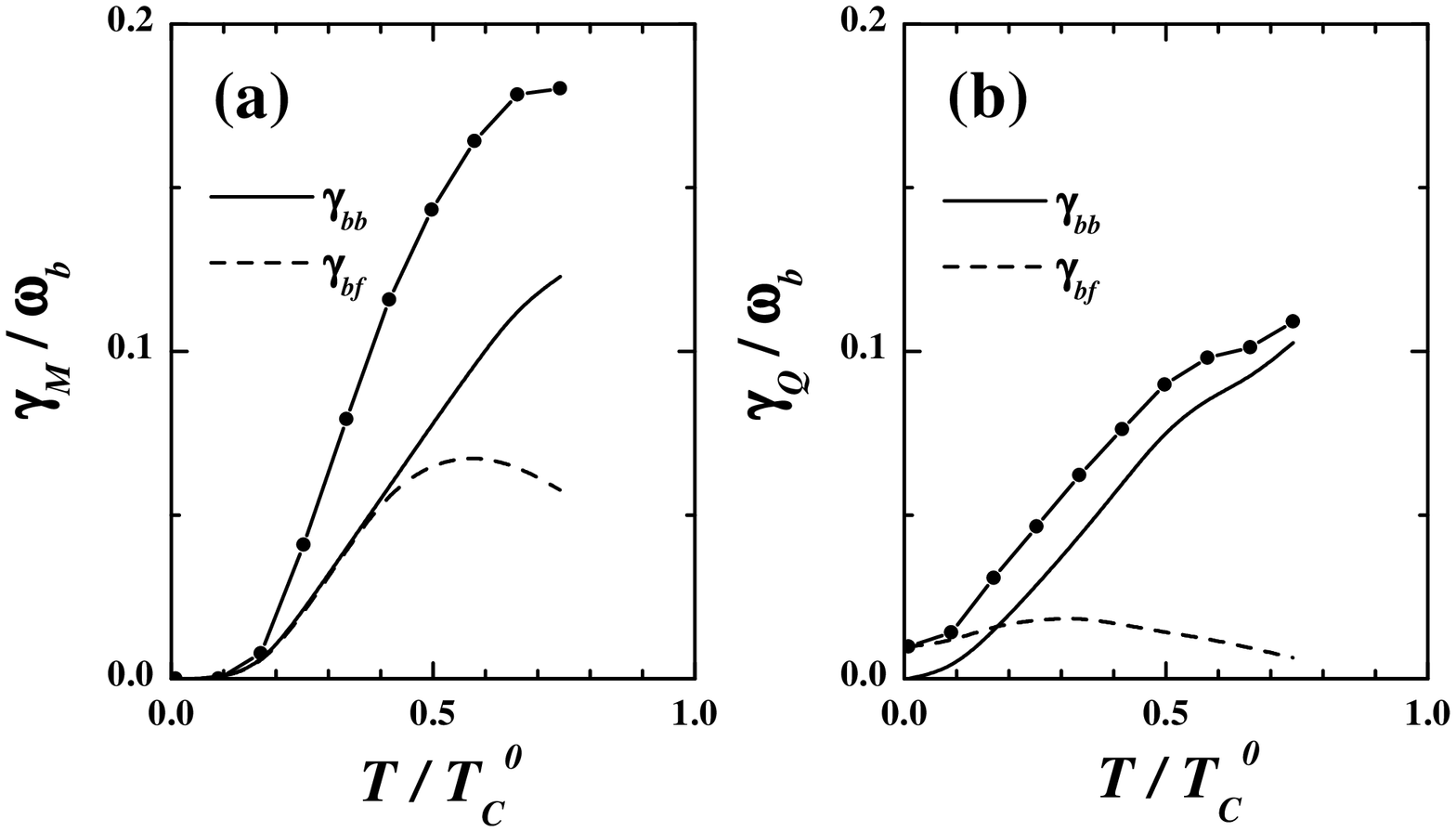}}
\caption{The same as in FIG. 5, but for the damping rates.}
\label{fig6}
\end{figure}

The last study in this subsection concerns the temperature
dependence of the eigenfrequency shifts and damping rates at a
specific Bose-Fermi coupling constant. In Figs. (5a) and (5b), we
report, respectively, our results for the monopole and quadrupole
mode frequencies as a function of the reduced temperature
$T/T_c^0$ at $\kappa =-2$. The corresponding damping rates are
shown in Figs. (6a) and (6b). The fulfillment of the generalized
Kohn theorem is checked in the inset of Fig. (5a), where the
calculated dipole frequency is very close to $\omega _b$ (or, more
precisely, $0.97\leq \omega _D/\omega _b\leq 1.0$) for all the
temperatures considered. At this specific value of $\kappa$, one
can see that both the monopole and quadrupole frequencies have a
downshift with increasing temperature, analogous to the results
obtained for a pure Bose gas \cite{giorgini,reidl}. In addition,
the behavior of the damping rates is also qualitatively similar
\cite{reidl,das,gp}.

\subsection{$^{87}{\rm Rb}-^{40}${\rm K}}

We now turn to consider a $^{87}{\rm Rb}-^{40}${\rm K} mixture composed of $%
2\times 10^4$ bosonic and $2\times 10^4$ fermionic atoms under the
conditions appropriate to the LENS experiments \cite{lens}. As in
experiment, we introduce the quantities $\alpha =m_f/m_b=0.463$ and $\beta =$
$\omega _f/\omega _b=1.47$ to parameterize the different mass and different
trapping frequency of the two species, which satisfy the constraint $\alpha
\beta ^2=1$ since both bosons and fermions experience the {\em same}
trapping potential. In addition, we take the $s$-wave Bose-Bose scattering
length $a_{bb}=99a_0=5.24$ {\rm nm} \cite{heinzen}, and fix the trapping
frequency $\omega _b=2\pi \times 91.7$ {\rm Hz}, which is the geometric
average of the axial and radial frequencies of Ref. \cite{lens}. The $s$%
-wave Bose-Fermi scattering length is varying, and in the experiment it can
be conveniently tuned by the Feshbach resonance \cite{simoni}. Notice that
the calculations presented here are restricted to the isotropic traps,
opposite to the cylindrical symmetric traps used in experiments. As a
result, our results are only useful in a qualitatively level.

\begin{figure}[tbp]
\centerline{\includegraphics[width=5.0cm,angle=-90,clip=]{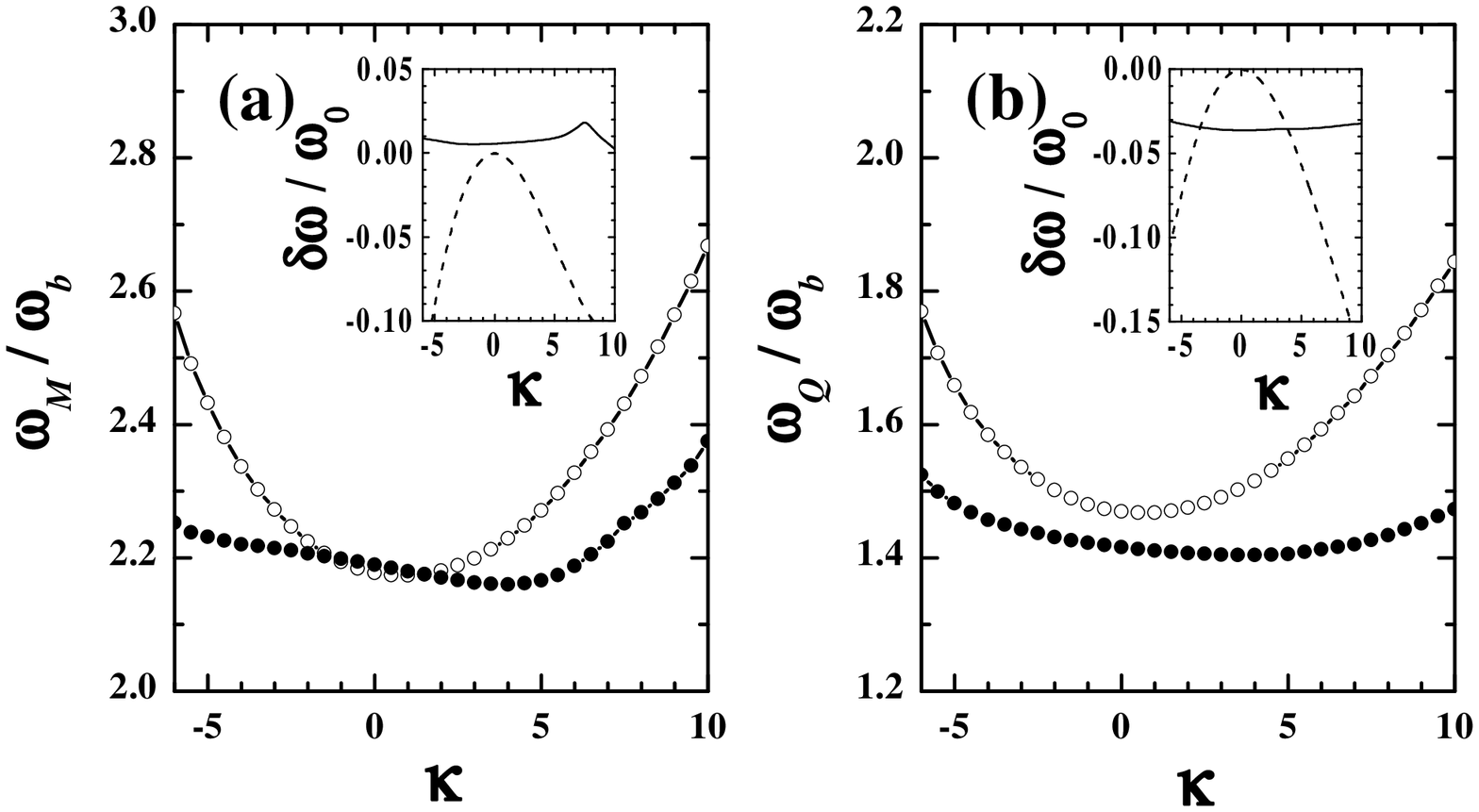}}
\caption{The dispersion relation of the monopole and quadrupole
condensate excitations for a mixture consisting of $2\times 10^4$ $^{87}{\rm Rb}$ and $%
2\times 10^4$ $^{40}${\rm K} atoms at $T=0.75T_c^0$, where
$T_c^0\approx 112$ {\rm nK}. The other parameters used in the
calculation are: $m_b=1.45\times
10^{-25}$ {\rm kg}, $\omega _b=2\pi \times 91.7$ {\rm Hz}, $m_f/m_b=0.463$, $%
\omega _f/\omega _b=1.47$, and $a_{bb}=99a_0=5.24$ {\rm nm}.}
\label{fig7}
\end{figure}

\begin{figure}[tbp]
\centerline{\includegraphics[width=5.0cm,angle=-90,clip=]{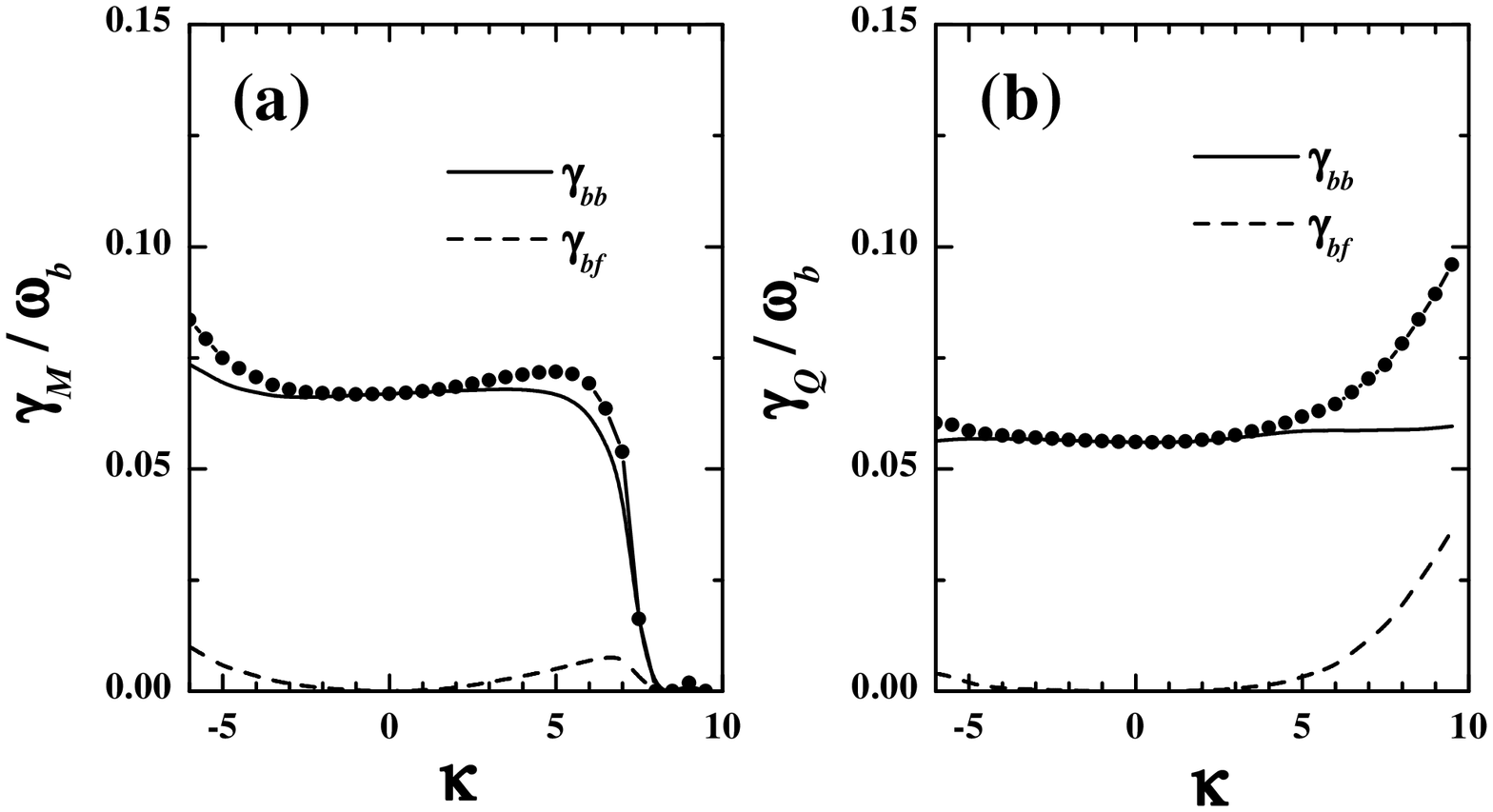}}
\caption{The same as in FIG. 7, but for the damping rates.}
\label{fig8}
\end{figure}

In Fig. 7, we plot the frequencies for the monopole and quadrupole
oscillations as a function of $\kappa $ at $T=0.75T_c^0$, where $%
T_c^0\approx 112$ {\rm nK}. Both the monopole and quadrupole frequencies
decrease slowly with increasing $\kappa $ up to $\kappa \approx 5$. Above
this value the frequencies gradually rise up. In the whole region of $\kappa
$, the variation of frequencies due to Bose-Fermi coupling is small.
However, it is still possible to be detected by the accurate frequency
measurement. For instance, at $\kappa =-6$, we find that the values of the
relative variation $\left( \omega _{\kappa =-6}-\omega _{\kappa =0}\right)
/\omega _{\kappa =0}$ for the monopole and quadrupole modes are,
respectively, $2.8\%$ and $7.6\%$, well within the experimental resolution.

The damping rate of the monopole and quadrupole modes at the same
temperature is shown in Figs (8a) and (8b), respectively. The behavior of
the damping rate against $\kappa $, that is, the decrease (increase) of the
monopole (quadrupole) damping rate across the demixing point $\kappa \approx
5$, is very similar to that in Figs. (3a) and (3b), except that the overall
magnitude is two times smaller. This behavior together with the slow rise up
of the mode frequency around $\kappa \approx 5$ thus may provide us a useful
signal to locate the onset of the phase transition of spatial separation.

\section{concluding remarks}

In this paper we have developed a theory for studying the
low-lying condensate oscillations of a spherically trapped
Bose-Fermi mixture at finite temperature in the collisionless
regime. In this theory, the unperturbed mode frequency is firstly
calculated within the static Hartree-Fock-Bogoliubov-Popov
approximation. The frequency correction, arising from the coupled
dynamics of the condensate, noncondensate, and degenerate Fermi
gas, is then taken into account perturbatively by means of the
random phase approximation. We have applied our theory to the
mixtures of $^{41}{\rm K}-^{40}${\rm K} and $^{87}{\rm
Rb}-^{40}${\rm K}, and have studied the dispersion relation of the
monopole and quadrupole condensate excitations as a function of
the Bose-Fermi coupling at various temperatures. The correctness
of our theory and numerical calculations is partly checked by the
fulfillment of the generalized Kohn theorem for the dipole
excitation. At a relatively high temperature we find that, as the
mixture moves towards demixing point with increasing Bose-Fermi
coupling, the damping rate of the monopole (quadrupole) excitation
increases (decreases). This behavior provides us a possible
signature to identify the phase transition of spatial separation.

\begin{acknowledgments}
We are very grateful to Dr. M. Modugno and Dr. G. Modugno for
simulating discussions. X.-J.L was supported by the K.C.Wong
Education Foundation, the Chinese Research Fund, and the NSF-China
under Grant No. 10205022.
\end{acknowledgments}


\begin{thebibliography}{99}
\bibitem{note4}  In other words, it arises from the process of one quantum
of the condensate oscillation $\omega $ being absorbed by a fermion below
the Fermi level with energy $\omega _i^F$, which is turned into a fermion
above the Fermi level with energy $\omega _j^F=\omega _i^F+\omega $.
\bibitem{note5}  That is, the divergence is removed by the substitutions, $%
\chi _{\tilde{m}\tilde{m}^{+}}\left( {\bf r},{\bf r}^{\prime };\omega
\right) \rightarrow \chi _{\tilde{m}\tilde{m}^{+}}\left( {\bf r},{\bf r}%
^{\prime };\omega \right) -\delta ({\bf r}-{\bf r}^{\prime })G_0^{irr}({\bf r%
})$, and $\chi _{\tilde{m}^{+}\tilde{m}}\left( {\bf r},{\bf r}^{\prime
};\omega \right) \rightarrow \chi _{\tilde{m}^{+}\tilde{m}}\left( {\bf r},%
{\bf r}^{\prime };\omega \right) -\delta ({\bf r}-{\bf r}^{\prime
})G_0^{irr}({\bf r})$. Here $G_0^{irr}({\bf r})$ is the singular part of the
ideal single-particle Green's function $G_0({\bf r+x}/2,{\bf r-x}/2;\omega
=0)$ that diverges as $1/x$ for $x$ $\rightarrow 0$ \cite{bruun}.
\end{thebibliography}
\end{document}